\documentclass{iaus}
\usepackage{graphicx,natbib}

\title[AGN in Optical Surveys] 
{A Complete Census of AGN and Their Hosts from Optical Surveys?}

\author[V. Wild et al.]  
{Vivienne Wild$^1$, Timothy Heckman$^2$, Paule Sonnentrucker$^2$,\\
  Brent Groves$^3$, Lee Armus$^4$, David Schiminovich$^5$,\\
 Benjamin Johnson$^6$, Lucimara Martins$^7$, \and Stephanie LaMassa$^2$}

\affiliation{$^1$ Institut d'Astrophysique de Paris, 75014 Paris,
  France.\\
Email: {\tt wild@iap.fr} \\[\affilskip]
$^2$ Dept. of Physics and Astronomy, Johns Hopkins University, Baltimore, MD 21218, USA \\[\affilskip]
$^3$ Leiden Observatory,  Leiden University, P.O. Box 9513, 2300 RA Leiden, The Netherlands \\[\affilskip]
$^4$ Caltech, Spitzer Science Center, MS 314-6, Pasadena, CA 91125, USA  \\[\affilskip]
$^5$ Dept. of Astronomy, Columbia University, NewYork, NY10027, USA \\[\affilskip]
$^6$ Institute of Astronomy, Madingley Road, Cambridge, CB3 0HA, UK
\\[\affilskip]
$^7$ NAT-Universidade Cruzeirodo Sul, Sao Paulo, Brazil \\
}

\pubyear{2010}
\volume{267}  
\pagerange{119--126}
\jname{Co-Evolution of Central Black Holes and Galaxies}
\editors{B.M.\ Peterson, R.S.\ Somerville, \& T.\ Storchi-Bergmann,
  eds.}

\def \Ledd {\ifmmode L_{\rm Edd} \else $L_{\rm Edd}$\fi}
\def \ledd {\ifmmode L_{\rm Edd} \else $L_{\rm Edd}$\fi}
\def \um {$\mu {\rm m}$}
\def  \ha {H$\alpha$}
\def  \hb {H$\beta$}

\bibpunct[, ]{(}{)}{;}{a}{}{,}

\def \mnras {\textit{MNRAS}}
\def \apj {\textit{ApJ}}

\def \apjl {\textit{ApJ}}

\begin{document}

\maketitle

\keywords{infrared: galaxies, galaxies: active, galaxies: starburst, 
galaxies: ISM}

\firstsection 
\section{Introduction}

In the last few decades, large optical spectroscopic surveys have
allowed enormous progress to be made in studying the galaxy
population, both the overall demographics and in the identification of
rare populations. The role of AGN on the evolution of the galaxy
population has also become an increasingly common topic of research,
largely driven by theoretical models and simulations which suggest
some additional input of energy into the ISM and IGM is required in
order to move galaxies from the blue to the red sequence and prevent
star formation in massive galaxies.

At low to intermediate redshifts, the sheer statistics of current
optical surveys allow us to identify and study significant numbers of
rare objects which represent transient phases in the lifetime of
galaxies: starburst, post-starburst and ``green-valley'' galaxies,
major mergers, and powerful AGN. These brief periods of major
disruption in the life of a galaxy may be insignificant compared to a
typical galaxy's total life span, but hugely significant globally,
i.e., when the effects are integrated over the whole population and the
age of the Universe. For example, summing the growth of black holes
attained during rare, high growth-rate phases (high $L/\ledd$) gives a
black hole mass density close to that observed in the local Universe,
implying that lower growth-rate events are relatively unimportant for
building black holes \citep{Yu:2002p3731,heckman04}.

Optical surveys allow us to locate sufficient samples of galaxies
experiencing transient phases of evolution. There is an obvious
benefit in using the information provided in the detection survey to
study the galaxies in greater detail, but, with the complex physical
processes at play during these periods of disruption, it is uncertain
how far optical studies alone can help us in understanding the true
impact of these transient phases on the life-cycle of the galaxy.  How
much physical information can we really infer from the optical when
faced with extreme star formation, dust production, AGN activity and
complicated spatial geometries of disturbed galaxies?

Here, we present ongoing work comparing star formation and AGN
properties derived from mid-infra-red spectra from the Spitzer space
telescope, to those derived from optical spectra. We combine
information from four surveys, three of which probe an ``unusual''
phase in the life of a galaxy and one control sample of ``ordinary''
galaxies.

\section{Samples}
{\underline {\it SSGSS}}: The control sample is the 
{\em Spitzer}--SDSS-{\em GALEX}
Spectroscopic Survey \citep[SSGSS, see][]{2009ApJ...705..885O}, a
representative sample of 100 local galaxies selected from the Sloan
Digital Sky Survey (SDSS) with {\em GALEX} photometry for follow-up
spectroscopy with the {\em Spitzer}--IRS. Their mass, color, star formation
rate and redshift distribution are representative of galaxies in the
SDSS survey.

{\underline {\it Dusty Balmer-strong}}: These galaxies were selected
from the SDSS-DR4 to be candidate ``post-starburst'' AGN with strong
Balmer absorption lines. But they also possess very high dust contents
which may mean the strong Balmer absorption lines are caused by dust
geometry rather than star formation history
\citep{2000ApJ...529..157P}.

{\underline {\it ULIRGs}}: These are the 10 ULIRGs in the IRAS Bright
Galaxy Sample; the {\em Spitzer} IRS low- and high-resolution observations
of these galaxies have been presented in detail in
\citet{Armus:2007p964}. Of the 10 BGS ULIRGs, six are found in the SDSS
spectroscopic sample, three do not lie within the SDSS footprint, and one is
too faint to have been targeted.

{\underline {\it Seyferts}}: The narrow-line Seyfert sample is composed
of the top 20 [O\,{\sc iii}]\,$\lambda 5007$ 
flux emitters out of all 
SDSS-DR4 main-sample
galaxies classified as Seyfert galaxies using the diagnostic line ratio
plot of [N\,{\sc ii}]/H$\alpha$ vs.\  [O\,{\sc iii}]/H$\beta$.

\begin{figure}[h!]
\begin{center}
 \includegraphics[width=6.5cm]{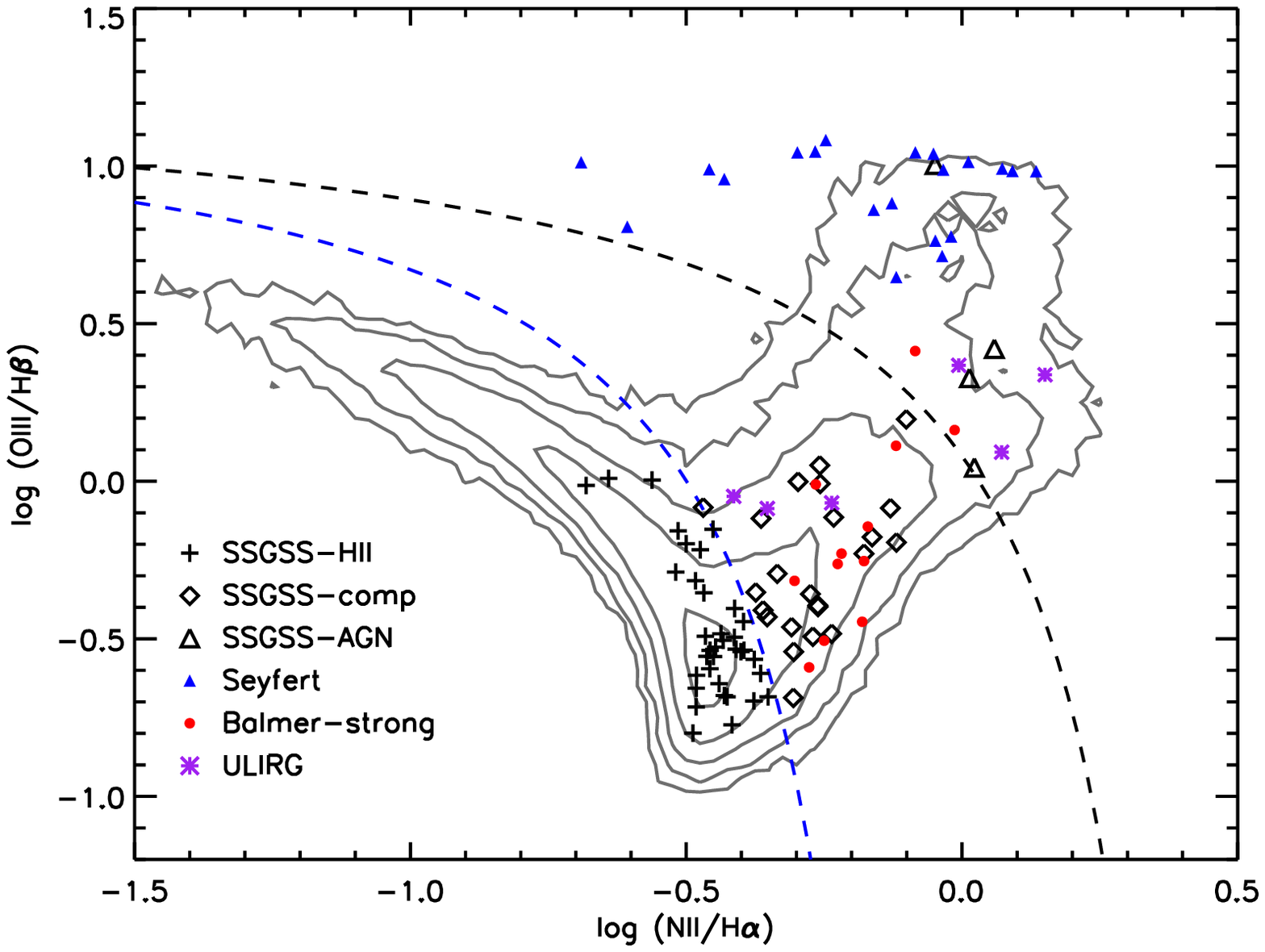} 
 \includegraphics[width=6.5cm]{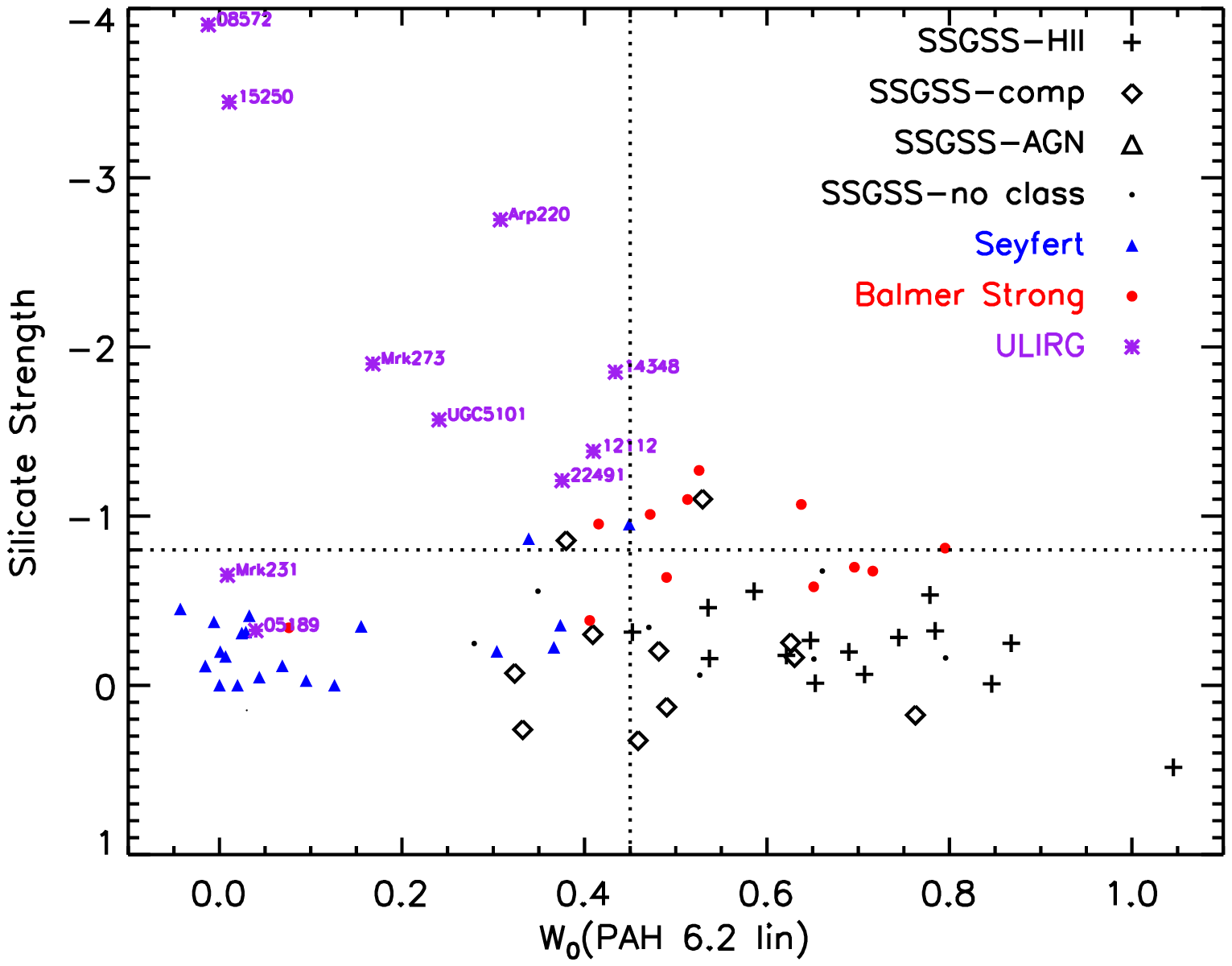} 
 \caption{{\it Left:} The optical BPT diagram of nebular emission-line
   ratios. Only those galaxies with signal-to-noise ratio
$S/N > $3 in all four emission lines are
   plotted. {\it Right:} The Spoon et al. (2007) IR diagnostic
   diagram. Only those galaxies with mean per-pixel
$S/N >$4 in the wavelength
   range 5.4--5.9\,\um\ are plotted. Formal upper limits on $W_{6.2}$ are not
   calculated: $W_{6.2}$ values $\lesssim0.05$ should be considered consistent
   with zero.}
   \label{fig:context}
\end{center}
\end{figure}

In Figure \ref{fig:context}, we place the four samples into context, for
both optical and IR astronomers. In the left-hand panel, we show the
BPT diagram of optical emission-line ratios which separates galaxies
with nebular emission dominated by star formation (left-hand branch)
from those dominated by higher ionization processes occurring in the
galactic nucleus (right-hand branch). The simplest explanation for the
population that lies between the two demarcation lines is that both
star formation and an AGN are present (``composite-AGN''). This is
supported by their younger mean stellar ages as measured from the
stellar continuum, compared to galaxies with ``pure-AGN'' emission
line ratios \citep{2003MNRAS.346.1055K,2006MNRAS.372..961K}. The gray
contours indicate the joint distribution of all SDSS-DR7 galaxies, the
symbols show the four samples described above. Both the dusty-Balmer
strong galaxies and the Seyferts were pre-selected to lie in the AGN
branch. All six ULIRGs, selected purely on IR luminosity, show optical
emission-line ratios which indicate the presence of an obscured
AGN. In the right-hand panel of Figure \ref{fig:context} we place all
our samples onto the diagram of \citet{2007ApJ...654L..49S} which
compares the rest frame equivalent width of the 6.2\,\um\ PAH feature
caused by on-going star formation, with the strength of the silicate
absorption feature at 10\,\um\ caused by dust absorbing the mid-IR
continuum along the line of sight (see following section). As expected
from their optical classification as AGN, both the Seyfert sample and
the SSGSS composite-AGN have on average lower $W_{6.2}$ than ordinary
star-forming galaxies, indicative of the destruction of small grain
PAHs in the strong radiation field of the QSO. The  Balmer-strong
AGN show stronger silicate absorption than ordinary galaxies, once
again qualitatively consistent with their large Balmer decrements
(H$\alpha$ to H$\beta$ emission-line ratios) in the optical.

\section{Dust}

The effect of dust grains intercepting stellar and nuclear light
complicates the analysis of integrated galaxy spectra. In the optical,
extinction of the light along the line-of-sight can be quantified
through a measured Balmer decrement (H$\alpha$ to H$\beta$ emission
line ratio), combined with an empirically measured dust extinction
curve (i.e. the Milky Way, LMC or SMC). However, the additional
scattered component that makes up a full {\it attenuation} curve
important for studies of integrated galaxy light has only been
measured empirically for the continua of starburst galaxies
\citep{1994ApJ...429..582C}. Emission lines are expected to suffer
more screenlike (less grey) attenuation than the optical continua, as
the emission must first pass through the dense birth clouds in which
the hot O and B stars that excite the nebular emission lines form
\citep{2000ApJ...539..718C,wild_psb,2008MNRAS.388.1595D}. An
additional complication arises in the case of composite galaxies, in
which the line emission originating from the narrow line region of the AGN
may suffer a different level and/or form of extinction compared to nebular
emission originating in star-forming regions.

Extinction in the mid-IR is dominated by features caused by the
stretching and bending modes of amorphous silicate grains. The
presence of silicate absorption can be used to infer the geometry of
material obscuring the light source
\citep{Levenson:2007p2766,Nenkova:2002p2744}.

\begin{figure}[h]
\begin{center}
 \includegraphics[width=7cm]{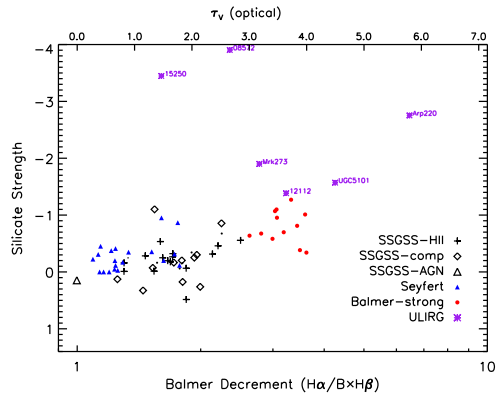} 
 \caption{Comparing dust obscuration in the IR (10\,\um\ silicate
   absorption) to dust obscuration in the optical (the Balmer
   decrement). We set $B=2.87$ (Case B recombination) for star forming
   and composite galaxies, and $B=3.1$ for pure-AGN. Only those
   galaxies with mean per-pixel $S/N >$4 in the wavelength range
   5.4--5.9\,\um, and SNR$S/N >$3 in their H$\alpha$ and H$\beta$ lines are
   plotted. The upper axis gives $\tau_V$, derived from the Balmer
   decrement assuming the two-component Charlot and Fall (2000) dust
   attenuation curve with $n_{\rm BC}=-1.3$ and $n_{\rm ISM}=-0.7$
   \citep{wild_psb}. }
   \label{fig:dust}
\end{center}
\end{figure}

In Figure \ref{fig:dust}, we compare two different measures of dust
content in the IR and optical regimes. Aside from the two extreme
ULIRGs, there is a general trend for the two measures of dust content
to track one another.  Although both axes describe dust content, the
trend is slightly surprising: any significant silicate absorption
applies optically thick clouds ($\tau_V>100$), which contrasts
strongly with the $\tau_V$ measured here in the optical. Therefore
these two quantities measure dust in entirely independent regions. It
is interesting to note that the two outliers from the trend (ULIRGs
15250 and 08572), with extremely strong silicate absorption implying
that the source of the mid-IR continuum is buried in a smooth,
geometrically and optically thick cloud, are composite-AGN in the
optical. Therefore some light from the central nucleus is escaping to
illuminate the narrow-line region, or the AGN is not the dominant
light source responsible for the mid-IR continuum and the dust is
not associated with the AGN torus.

\section{Low-Ionization Emission Lines}

In star-forming galaxies, the dominant excitation mechanism for nebular
emission lines is UV radiation from young, hot stars.  When star
formation is the dominant mechanism of excitation, \ha\ is one of the
best calibrated measures of instantaneous star formation rate in
galaxies. However, because the \ha\ line lies in the optical, to
obtain accurate SFR estimates we must correct for dust attenuation. In
this section we address the question of how accurate the standard
methods of correction for dust attenuation in the optical are, by
comparing the \ha\ luminosity to the luminosity of the lower
ionization neon lines in the mid-IR. To first order, the strengths of
these lines are expected to be strongly correlated, especially in the
relatively massive, and therefore metal rich, galaxies that make up
the majority of our samples.

\begin{figure}[h]
\begin{center}
 \includegraphics[width=12cm]{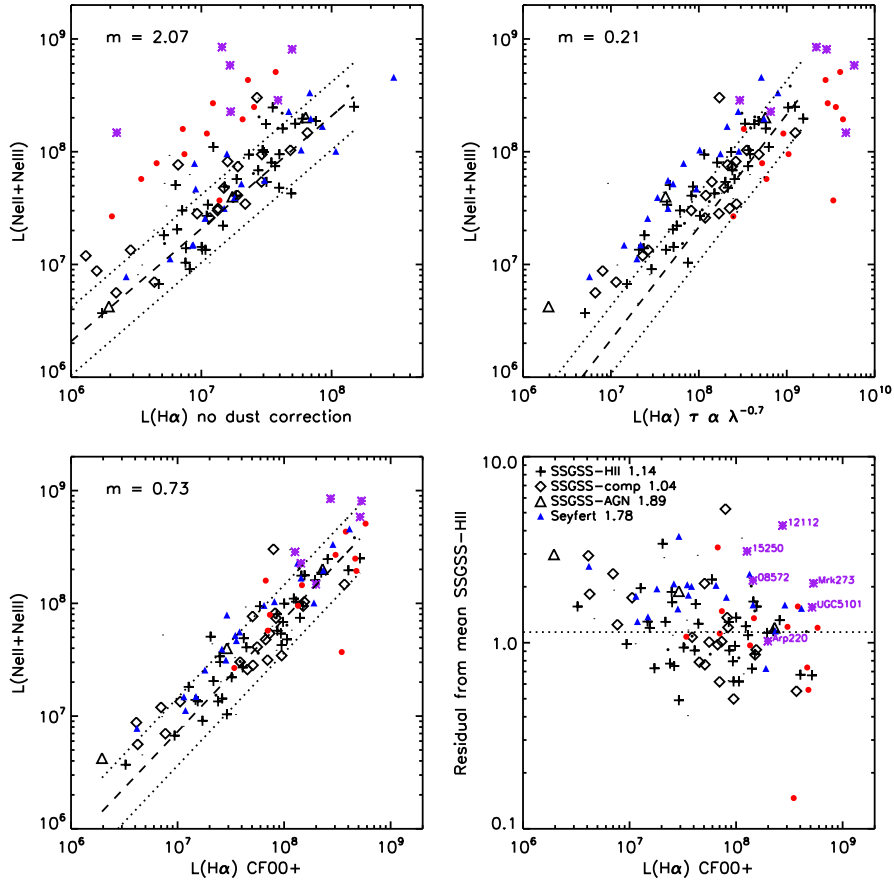} 
 \caption{The luminosity of \ha\ in the optical vs.\ the low
   ionization Ne lines in the mid-IR. {\it Top left:} \ha\ is not
   corrected for dust attenuation. {\it Top right:} \ha\ is
 corrected for dust attenuation using a single component CF00
 model. {\it Bottom left:} \ha\ is corrected for dust attenuation
 using a double component CF00 model. In all three panels, the dashed
 line shows the best-fit linear fit to the SSGSS star-forming
 galaxies, and the dotted lines show a factor of two above and
 below. {\it Bottom right:} The residual from the linear fit in the
 bottom-left panel.}
   \label{fig:neha}
\end{center}
\end{figure}

In Figure \ref{fig:neha}, we plot \ha\ vs.\ Ne luminosities for all of
the samples, firstly with \ha\ uncorrected for dust attenuation, then
corrected with two different dust attenuation laws used in the
literature. In the top-left panel, where no correction for dust
attenuation has been applied to the \ha\ luminosities, we immediately
notice that the Balmer-strong galaxies and ULIRGs lie above the SSGSS
galaxies as expected for their large dust contents. In the top-right
panel, the dust law is a single component model with $\tau \propto
\lambda^{-0.7}$. This law provides a good fit to the UV and IR
continua of starburst galaxies \citep[][hereafter
CF00]{2000ApJ...539..718C}, but we see that it does not provide a good
fit to the nebular emission lines, causing considerable overcorrection
of the optical emission lines in the dustiest galaxies, and
undercorrection in the least dusty Seyferts compared to normal
SSGSS--H\,{\sc ii} galaxies. In the bottom-left panel, we employ the full
two-component dust model of CF00, where the second term allows for the
additional, greyer, attenuation of the stellar birth clouds with an
exponent of $-1.3$ \citep{wild_psb,2008MNRAS.388.1595D}. There is a
noticeable improvement in the scatter. In each panel, the dashed line
shows a linear least-squares fit to the SSGSS--H\,{\sc ii} galaxies, with the
dotted lines indicating a factor of two above and below the
best-fit. In the bottom-right we plot the residual away from this
best-fit line using the two-component CF00 model. The scatter is
around a factor of two, even for the extreme dusty Balmer-strong
galaxies. Even the ULIRGs do not show a dramatic underprediction of
their \ha\ luminosities, despite their optical Balmer decrements
indicating as much as six magnitudes of extinction and their silicate
absorption indicating hundreds of magnitudes. We note a slight mean
offset for the AGN, in the sense that they have stronger 
Ne\,{\sc ii}+Ne\,{\sc iii}
emission for their \ha\ luminosities. This could be due to the harder
ionizing radiation field of the AGN, or that the true dust attenuation
law is slightly different in the narrow line region of the
AGN. Although the effect is not large, further study may be able to
untangle these two effects.

\section{High-Ionization Emission Lines}

Given the success of the 2 component dust law for correcting \ha\ for
dust attenuation in even the dustiest galaxies in the local Universe,
we now turn to the question of using [O\,{\sc iii}] luminosity as a proxy for
black hole accretion rate. Although originating from the narrow line
region, far removed spatially from the central nucleus with its dense
obscuring torus, the [O\,{\sc iii}] emission must still suffer extinction from
the host galaxy, and possibly also from dust within the narrow line
region. This adds an additional degree of complexity to the problem:
the \ha\ and \hb\ emission lines with which we calculate the magnitude
of the dust attenuation in the optical, may originate at least in part
from star-forming regions completely spatially separated from the
narrow-line region in which we are interested. However, given the
small residual offset between Ne and \ha\ emission presented in the
previous section, even for the strongest AGN in the SDSS (the Seyfert
sample), this effect cannot be large.

\begin{figure}[h]
\begin{center}
\includegraphics[width=6.5cm]{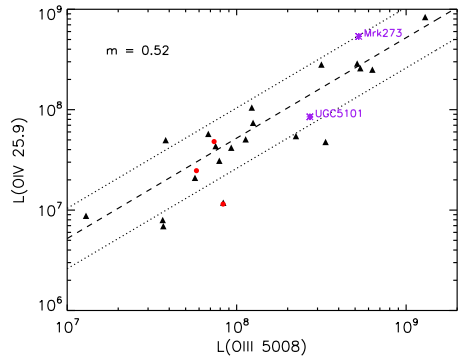}
\end{center}
\caption{Dust-corrected $L$([O\,{\sc iii}]) vs.\ $L$([O\,{\sc iv}]) 
for the Seyfert sample
 (filled triangles), the three Balmer-strong galaxies with [O\,{\sc iv}] detections
  (red circles) and the two ULIRGs with both [O\,{\sc iv}] detections and SDSS
  spectra.  }
\label{fig:oiiioiv}
\end{figure}

In Figure \ref{fig:oiiioiv}, we plot individual detections of [O\,{\sc
iv}]\,$\lambda$25.9\,\um\ versus [O\,{\sc iii}]\,$\lambda 5007$, using
the Balmer decrement and two component CF00 model to correct [O\,{\sc
iii}] for dust attenuation. The mid-IR [O\,{\sc iv}] line lies too
close to an Fe\,{\sc ii} line for deconvolution in the low resolution
{\em Spitzer} spectra. Therefore we only plot the the Seyfert, ULIRG
and dusty Balmer-strong galaxy samples, for which we have high
resolution spectra. Of the galaxies in this figure, two of the
Balmer-strong galaxies and both of the ULIRGs are optically classified
as pure-AGN, and the majority of the [O\,{\sc iii}] is therefore
likely to originate from the AGN rather than from star formation. For
the third Balmer-strong galaxy, a small correction has been applied to
account for the star formation contamination of the [O\,{\sc iii}]
line. At higher ionization, the [O\,{\sc iv}] line is expected to
suffer less contamination from star formation.

Once again, we find that where there are individual detections of both
[O\,{\sc iv}] and [O\,{\sc iii}], the line luminosities are strongly
correlated after appropriate correction for dust attenuation of
[O\,{\sc iii}], even in extremely dusty galaxies. Although further
work is required to investigate those galaxies where [O\,{\sc iv}] is
not detected in the {\em Spitzer} spectra, this preliminary result
suggests that dust attenuation is not a significant cause for concern
in the use of optical spectra for deriving black hole accretion rates.

\section{Discussion}

In this preliminary investigation comparing the mid-IR and optical
emission lines of three samples of unusual galaxies in the local
Universe, and a sample of more ordinary galaxies, we find little
evidence that the optical wavelength regime, used alone, would provide
a misleading picture in terms of either star formation rates or black
hole accretion rates. Each of the three samples is extreme in some
way: the Seyferts are the strongest [O\,{\sc iii}] emitters in the local
Universe, the Balmer-strong galaxies are some of the dustiest AGN in
the SDSS, the ULIRGs are unusually bright in the IR, implying
substantial dust contents. We have shown how each of these samples
fits into the global galaxy population, both in the mid-IR and in the
optical.

It is perhaps surprising that, given the geometrically complicated
dust distributions of some galaxies, and the spatial separation
between the narrow line region and the star-forming regions in
composite-AGN, that nebular emission lines in the optical wavelength
regime can be adequately corrected for the effects of dust
attenuation. With further investigation of these samples, we hope to
understand more about the limitations, or otherwise, of the optical
wavelength regime for studying unusual galaxies.


\end{document}